\documentclass[twocolumn]{autart}    

\usepackage{graphicx}          
\usepackage{color}

\usepackage{amsmath}

\usepackage{amssymb}

\def \be {\begin{equation}}

\def \ee {\end{equation}}

\def \nn {\nonumber}

\begin{document}

\begin{frontmatter}

\title{A Differential-Cascaded Approach for Adaptive Control of Robot Manipulators} 

\author{Hanlei Wang}\ead{hlwang.bice@gmail.com}

\address{Science and Technology on Space Intelligent Control Laboratory, Beijing Institute of Control Engineering, Beijing 100094, China}


\begin{abstract}                          
This paper investigates adaptive control of nonlinear robot manipulators with parametric uncertainty. Motivated by generating closed-loop robot dynamics with enhanced transmission capability of a reference torque and with connection to linear dynamics, we develop a new adaptive approach by exploiting forwardstepping design and inertia invariance, yielding differential-cascaded closed-loop dynamics. With this approach, we propose a new class of adaptive controllers for nonlinear robot manipulators. Our particular study concerning adaptive control of robots exhibits a design methodology towards establishing the connection between adaptive control of highly nonlinear uncertain systems (e.g., with a variable inertia matrix) and linear dynamics (typically with the same or increased order), which is a long-standing intractable issue in the literature.
\end{abstract}

\end{frontmatter}

\section{Introduction}

The study of the control problem for robot manipulators over the past several decades has generated a variety of interesting results (see, e.g., \cite{Takegaki1981_ASME,Craig1987_IJRR,Slotine1987_IJRR,Middleton1988_SCL,Slotine1989_AUT,Li1989_SCL,Ortega1989_AUT,Cheah2003_TRA,Cheah2006_IJRR,Dixon2007_TAC,Wang2017_TAC}).
The fundamental result in \cite{Slotine1987_IJRR} is generally considered to be a standard for accommodating the nonlinearity and uncertainty of robot manipulators, and this adaptive controller is also referred to as a passivity-based controller \cite{Ortega1989_AUT}. This kind of passivity-based design methodology has been extensively studied in many other contexts such as observer design for nonlinear robots \cite{Berghuis1993_TRA}, adaptive attitude control of spacecraft \cite{Slotine1990_TAC,Egeland1994_TAC}, and adaptive image-space control of robots \cite{Liu2006_TRO,Wang2015_AUT,Li2021_TCST}. Other results beyond the passivity-based approach appear in, e.g., \cite{Seo2009_SCL,Seo2016_CDC,Yang2017_JGCD,Xia2020_IJACSP}, which can be categorized as the non-certainty equivalent approach or immersion and invariance approach \cite{Astolfi2003_TAC} and typically involve either the filtering of the tracking errors and the control input or dynamic scaling (for guaranteeing the solvability of a partial differential equation).

The distinctive properties of the adaptive approach in \cite{Slotine1987_IJRR} are that neither the joint acceleration nor the inversion or invertibility of the inertia matrix is required and that the closed-loop dynamics are still nonlinear and physical (even in the context with known parameters), in contrast to the adaptive inverse dynamics approaches (see, e.g., \cite{Craig1987_IJRR,Middleton1988_SCL,Li1989_SCL}). This also leads to an important tendency concerning adaptive control of robot manipulators, i.e., the connection between adaptive control of nonlinear robots and linear dynamics has almost been abandoned ever since the result in \cite{Slotine1987_IJRR}. The unfavorable aspect of this tendency is reflected in, e.g., control of robot manipulators involving interaction (for instance, bilateral teleoperation or teaching operation), and quantification of the performance of the closed-loop robot dynamics (see, e.g., \cite{Spong1990_TAC}). In particular, in bilateral control of teleoperators, the control for the master and slave robots is expected to ensure the transmission performance from a human operator to the environment or to the consensus dynamics (see, e.g., \cite{Wang2020b_AUT}). This fundamentally relies on establishing the connection between control of robot manipulators with respect to a reference torque and linear dynamics (differing from the Slotine and Li approach \cite{Slotine1987_IJRR} and also from the adaptive inverse dynamics approach \cite{Craig1987_IJRR,Middleton1988_SCL,Li1989_SCL}), which is a long-standing intractable problem in the literature. The nonlinear dynamics with uncertain variable inertia matrix yield the challenge of achieving the transmission of a reference torque across this dynamics (we may note that under the standard (adaptive) inverse dynamics control, the closed-loop dynamics governing the reference torque and the position tracking error are not linear due to the configuration-dependent inertia matrix).

In this paper, we propose a new design methodology for adaptive control of robot manipulators to establish the connection between controlled robot dynamics and linear dynamics. The proposed approach relies on the exploitation of inertia invariance and forwardstepping design of reference dynamics (see \cite{Wang2019_ACC,Wang2020_arXiv}). The new class of adaptive controllers, derived by using forwardstepping concerning the reference dynamics with inertia invariance, is claimed as achieving ``controlled physics'' in the sense that the unknown inertia matrix of the robot manipulator is retained as the equivalent inertia of the closed-loop dynamics. As has been and will be witnessed, the inertia invariance formally referred to here may be the most essential and physical nature for adaptive control of robot manipulators (without involving inversion of the estimated inertia matrix and joint acceleration measurement); in this sense, the adaptive controllers in \cite{Slotine1987_IJRR,Slotine1989_AUT} and the non-certainty equivalent approach \cite{Seo2009_SCL} can be considered as some examples of this nature, and there are a variety of adaptive controllers beyond these particular examples, as is shown in our result. The formulation of forwardstepping adaptive control with inertia invariance, from a general perspective of nonlinear control, establishes the connection between adaptive control of nonlinear robot manipulators and linear control, with an interesting feature analogues to the standard Taylor's theorem in calculus concerning approximation of functions. In particular, the closed-loop dynamics under forwardstepping control with inertia invariance approach linear dynamics with a remainder in the sense of certainty equivalence; the remainder can be considered as quantification of the approximation of linear dynamics to the closed-loop robot dynamics. The order increment of the remainder, implying that the closed-loop robot dynamics tend to be closer to linear dynamics, can be achieved by specifying reference dynamics with an increment of their order.

\section{Equations of Motion of Robot Manipulators}

The equations of motion of a robot manipulator can be written as \cite{Slotine1991_Book,Spong2006_Book}
\be
\label{eq:1}
M(q)\ddot q+C(q,\dot q)\dot q+g(q)=\tau+\tau^\ast
\ee
where $q\in R^n$ is the joint position, $M(q)\in R^{n\times n}$ is the inertia matrix, $C(q,\dot q)\in R^{n\times n}$ is the Coriolis and centrifugal matrix, $g(q)\in R^n$ is the gravitational torque, $\tau\in R^n$ is the joint control torque, and $\tau^\ast\in R^n$ is the reference torque input (e.g., the interaction torque between the robot and a human operator, or the disturbance). To facilitate the reference in the sequel, two standard properties associated with the dynamics (\ref{eq:1}) are formulated as follows.

\emph{Property 1 (\cite{Slotine1991_Book,Spong2006_Book}):} The inertia matrix $M(q)$ is symmetric and uniformly positive definite.

\emph{Property 2 (\cite{Slotine1991_Book,Spong2006_Book}):} The dynamics (\ref{eq:1}) define a linear mapping of a constant parameter vector $\vartheta$.

\section{Adaptive Control}

 Let $q_d\in R^n$ denote the desired joint position with $q_d, \dot q_d, \dots, d^{\ell+1} q_d/d t^{\ell+1}$ being bounded, $\ell=1,2,\dots$. The control objective is to ensure that the position tracking error $\Delta q=q-q_d$ converges to zero as $t\to\infty$ with $\tau^\ast=0$, and that the transmission performance from $\tau^\ast$ to $\Delta q$ can be quantified by some performance index.
\subsection{Adaptive Control with Differential-Cascaded-Degree One}

Define a vector $z\in R^n$ with reference dynamics given as
\begin{align}
\label{eq:2}
\dot z=\ddot q_d-\alpha_1\Delta \dot q-\alpha_0\Delta q+\lambda_{\mathcal S}^\ast\lambda_c \hat M(q)(\dot q-z)
\end{align}
where $\alpha_0$, $\alpha_1$, and $\lambda_c$ are positive design constants, and $\lambda_{\mathcal S}^\ast$ is a design constant (scaling). Define
\be
\label{eq:3}
s=\dot q-z
\ee
and a regressor matrix $Y$ as (due to Property 2)
\be
\label{eq:4}
 M(q)\dot z+ C(q,\dot q)\dot q+ g(q)- {\dot M}(q)s-\lambda_c  M(q)s=Y(q,\dot q,z,\dot z)\vartheta
.\ee
The control torque is given as
\begin{align}
\label{eq:5}
\tau=&-\lambda_c \hat M(q) s+\hat M(q)\dot z+\hat C(q,\dot q)\dot q\nn\\
&+\hat g(q)-\hat {\dot M}(q)s+W\dot{\hat \vartheta}
\end{align}
where $\hat\vartheta$ is the estimate of $\vartheta$, and $W$ is a matrix defined by the dynamics
\be
\label{eq:6}
\dot W=-\lambda_c W+Y(q,\dot q,z, \dot z).
\ee
The adaptation law for $\hat \vartheta$ is given as
\be
\label{eq:7}
\dot{\hat \vartheta}=-\Gamma W^T s
\ee
where $\Gamma$ is a symmetric positive definite matrix.

The closed-loop dynamics of the manipulator can be described by a differential-cascaded system as
\be
\label{eq:8}
\begin{cases}
\Delta\ddot q=-\alpha_1\Delta \dot q-\alpha_0\Delta q+\lambda_{\mathcal S}^\ast\lambda_c \hat M(q) s+\dot s\\
\frac{d}{dt}[M(q)s]=-\lambda_c M(q)s+\frac{d}{dt}[W\Delta \vartheta]+\lambda_c W\Delta\vartheta+\tau^\ast\\
\dot {\hat\vartheta}=-\Gamma W^T s
\end{cases}
\ee
where $\Delta \vartheta=\hat\vartheta-\vartheta$. We observe that the lower two subsystems of (\ref{eq:8}) are also differential-cascaded in the sense that the interconnection component $\frac{d}{dt}[W\Delta \vartheta]+\lambda_c W\Delta\vartheta$ involves the derivative of the parameter estimate $\hat\vartheta$.
With degree reduction \cite{Wang2019_ACC} concerning $\frac{d}{dt}[W\Delta\vartheta]$, the closed-loop dynamics of the manipulator can further be written as
\be
\label{eq:9}
\begin{cases}
\Delta\ddot q=-\alpha_1\Delta \dot q-\alpha_0\Delta q+\lambda_{\mathcal S}^\ast \lambda_c\hat M(q) s+\dot s\\
\frac{d}{dt}[M(q)s-W\Delta \vartheta]=-\lambda_c[M(q)s-W\Delta \vartheta]+\tau^\ast\\
\dot {\hat\vartheta}=-\Gamma W^T s.
\end{cases}
\ee

\emph{Theorem 1:} The adaptive controller given as (\ref{eq:5}) and (\ref{eq:7}) with $z$ given as (\ref{eq:2}) ensures the convergence of the tracking errors, i.e., $\Delta q\to 0$ and $\Delta \dot q\to 0$ as $t\to\infty$ with $\tau^\ast=0$. The quantitative nonlinearity measure of the dynamics of the transmission from $\tau^\ast$ to $\Delta q$ (with respect to linear dynamics) in the sense of certainty equivalence can be specified by the residual error $\dot s$.

\emph{Proof:} The application of the standard linear system theory to the second subsystem of (\ref{eq:9}) immediately yields the conclusion that $M(q)s-W\Delta\vartheta\in\mathcal L_2\cap\mathcal L_\infty$. Let $\psi=M(q)s-W\Delta\vartheta$, and we have that
\be
\label{eq:10}
M(q)s=W\Delta \vartheta+\psi.
\ee
Due to the result that $\psi\in\mathcal L_2$ and Property 1, there exists a positive constant $\ell^\ast$ such that $\int_0^t\psi^T M^{-1}(q)\psi d\sigma\le \ell^\ast$, $\forall t\ge 0$. Consider the quasi-Lyapunov function candidate $V=(1/2)\int_0^t s^T M(q)sd\sigma+(1/2)\Delta \vartheta^T \Gamma^{-1}\Delta \vartheta+[\ell^\ast-\int_0^t\psi^T M^{-1}(q)\psi d\sigma]$ with the last term of $V$ being chosen in accordance with the standard practice (see, e.g., \cite[p.~118]{Lozano2000_Book}), and we have that $\dot V\le-(1/4)s^T M(q)s\le 0$ where we have used the following result from the standard basic inequalities $$s^T\psi\le (1/4)s^TM(q)s+\psi^T M^{-1}(q)\psi.$$ This implies that $s\in\mathcal L_2$ and $\hat\vartheta\in\mathcal L_\infty$. Therefore, $\hat M(q)$ is bounded. From (\ref{eq:10}), we have that $W\Delta \vartheta\in\mathcal L_2$. The dynamics (\ref{eq:9}) can be further written as
\be
\label{eq:11}
\begin{cases}
\Delta\ddot q=-\alpha_1\Delta \dot q-\alpha_0\Delta q+\lambda_{\mathcal S}^\ast\lambda_c [M(q) s-W\Delta\vartheta]\\
\qquad+\lambda_{\mathcal S}^\ast\lambda_c\{[\hat M(q)-M(q)]s+ W\Delta\vartheta\}+\dot s\\
\frac{d}{dt}[M(q)s-W\Delta \vartheta]=-\lambda_c[M(q)s-W\Delta \vartheta]+\tau^\ast\\
\dot {\hat\vartheta}=-\Gamma W^T s.
\end{cases}
\ee
Consider the first subsystem of (\ref{eq:11}) with $\lambda_{\mathcal S}^\ast \lambda_c[M(q) s-W\Delta\vartheta]$, $\lambda_{\mathcal S}^\ast\lambda_c\{[\hat M(q)-M(q)]s+ W\Delta\vartheta\}$, and $\dot s$ as the inputs, respectively and $x=[\Delta q^T,\Delta \dot q]^T$ as the output. From the standard linear system theory, the first subsystem of (\ref{eq:11}) with the three inputs being zero is exponentially stable and strictly proper. From the input-output properties of exponentially stable and strictly proper linear systems \cite[p.~59]{Desoer1975_Book}, the input $\lambda_{\mathcal S}^\ast \lambda_c[M(q) s-W\Delta\vartheta]$ yields an output that is square-integrable and bounded, and the input $\lambda_{\mathcal S}^\ast\lambda_c\{[\hat M(q)-M(q)]s+ W\Delta\vartheta\}$ yields an output that is also square-integrable and bounded. From \cite{Wang2020_AUT}, we have that the input $\dot s$ yields an output that is square-integrable. Hence, the application of the standard superposition principle for linear systems, we have that $x\in\mathcal L_2$, and therefore $\Delta q\in\mathcal L_2$ and $\Delta \dot q\in\mathcal L_2$. This implies that $\Delta q\in\mathcal L_\infty$ and $\Delta q\to 0$ as $t\to \infty$ from the standard properties of functions (see, e.g., \cite{Lozano2000_Book}). The boundedness of $\dot q$ and $\ddot q$ can be demonstrated in a similar way as \cite{Middleton1988_SCL}. Specifically, the result that $\Delta \dot q\in \mathcal L_2$ implies that $\dot q$ can be written as the sum of a square-integrable variable and a bounded variable. From (\ref{eq:2}), we have that $\dot z$ can be expressed as the sum of a square-integrable variable and a bounded variable.  Using (\ref{eq:3}) yields the result that $z=\dot q-s$, and this implies that $z$ can be considered to be the sum of a square-integrable variable and a bounded variable. The subsequent analysis can be completed via following similar practice as in \cite{Middleton1988_SCL}.

As the estimated parameter $\hat\vartheta$ converges to $\vartheta$ and without considering $\dot s$ in the first subsystem of (\ref{eq:8}), the mapping from $\tau^\ast$ to $\Delta q$ is governed by a third-order linear system (the composite of linear parts in the second-order reference dynamics and closed-loop robot dynamics). Hence, in the sense of certainty equivalence, the closed-loop robot dynamics approximately behaves as third-order linear dynamics with a remainder $\dot s$. \hfill \text{\small $\blacksquare$}

\subsection{Adaptive Control with Differential-Cascaded-Degree Two}

For this case,  the vector $z$ is defined by reference dynamics given as
\begin{align}
\label{eq:12}
\ddot z=&\dddot q_d-\alpha_2\Delta \ddot q-\alpha_1\Delta \dot q-\alpha_0\Delta q\nn\\
&+\lambda_{\mathcal S}^\ast\frac{d}{dt}[\hat M(q) (\dot q-z)]+\lambda_c\lambda_{\mathcal S}^\ast \hat M(q) (\dot q-z)
\end{align}
where $\alpha_0$, $\alpha_1$, and $\alpha_2$ are positive design constants that are chosen such that the polynomial $\theta^3+\alpha_2\theta^2+\alpha_1\theta+\alpha_0$ with $\theta$ being the free variable is a Hurwitz polynomial. With the adaptive controller given as (\ref{eq:5}) and (\ref{eq:7}), the closed-loop dynamics can be described by the following differential-cascaded system
\be
\label{eq:13}
\begin{cases}
\Delta\dddot q=-\alpha_2\Delta\ddot q-\alpha_1\Delta \dot q-\alpha_0\Delta q\\
\qquad+\lambda_{\mathcal S}^\ast \lambda_c\hat M(q) s+\lambda_{\mathcal S}^\ast\frac{d}{dt}[\hat M(q)s]+\ddot s\\
\frac{d}{dt}[M(q)s-W\Delta \vartheta]=-\lambda_c[M(q)s-W\Delta \vartheta]+\tau^\ast\\
\dot {\hat\vartheta}=-\Gamma W^T s.
\end{cases}
\ee

\emph{Theorem 2:} The adaptive controller given as (\ref{eq:5}) and (\ref{eq:7}) with $z$ given as (\ref{eq:12}) ensures the convergence of the tracking errors, i.e., $\Delta q\to 0$ and $\Delta \dot q\to 0$ as $t\to\infty$ with $\tau^\ast=0$. The quantitative nonlinearity measure of the dynamics of the transmission from $\tau^\ast$ to $\Delta q$ (with respect to linear dynamics) in the sense of certainty equivalence can be specified by the residual error $\ddot s$.

Using degree reduction to the first subsystem of (\ref{eq:13}) yields (similar to \cite{Wang2019_ACC})
 \begin{align}
\begin{cases}
\Delta \dot q=\Delta \dot q\\
\Delta \ddot q=(\Delta \ddot q-\dot s)+\dot s\\
\frac{d}{dt}(\Delta \ddot q-\dot s)=-\alpha_2(\Delta \ddot q-\dot s)-\alpha_1\Delta \dot q-\alpha_0\Delta q\\
\qquad\qquad-\alpha_2 \dot s+\lambda_{\mathcal S}^\ast \lambda_c\hat M(q) s+\lambda_{\mathcal S}^\ast\frac{d}{dt}[\hat M(q)s]
.\end{cases}
\end{align}
With $x=[\Delta q^T,\Delta \dot q^T,(\Delta \ddot q-\dot s)^T]^T$ as the state, the proof of Theorem 2 can then be performed with similar procedures as in that of Theorem 1.

\emph{Remark 1:} In comparison with (\ref{eq:8}), the differential-cascaded dynamics (\ref{eq:13}) involve a closed-loop robot system that can be approximated more accurately by third-order linear dynamics due to the order increment of the remainder [namely the remainder $\ddot s$ as compared with $\dot s$ in (\ref{eq:8})]. Depending on the specific context, we can implement appropriate adjustment of the first subsystem of (\ref{eq:13}) by redesigning the reference dynamics, e.g., redesigning the interconnection component. For instance, if we expect that the linear part $\Delta \dddot q=-\alpha_2\Delta \ddot q-\alpha_1\Delta \dot q-\alpha_0\Delta q$ approximates better to the reference torque $\tau^\ast$, we can redefine the reference dynamics such that
\begin{align}
\Delta\dddot q=&-\alpha_2\Delta\ddot q-\alpha_1\Delta \dot q-\alpha_0\Delta q\nn\\
&+\lambda_{\mathcal S}^\ast \lambda_c\hat M(q) s+\lambda_{\mathcal S}^\ast\frac{d}{dt}[\hat M(q)s+\lambda_c \hat M(q)s]+\ddot s
\end{align}
which incorporates the low-frequency part of the derivative of $\tau^\ast$.

\subsection{Adaptive Control with Arbitrary Differential-Cascaded-Degree}

Consider a definition of reference dynamics given as
\begin{align}
\label{eq:16}
\frac{d^\ell z}{dt^\ell}=&\frac{d^{\ell+1}q_d}{dt^{\ell+1}}-\alpha_\ell \frac{d^\ell \Delta q}{dt^\ell}-\dots-\alpha_0\Delta q\nn\\
&+\lambda_{\mathcal S}^\ast\frac{d}{dt}[\hat M(q) (\dot q-z)]+\lambda_c\lambda_{\mathcal S}^\ast \hat M(q) (\dot q-z)
\end{align}
where $\alpha_0,\dots,\alpha_\ell$ are positive constants that are chosen such that $\theta^{\ell+1}+\alpha_\ell \theta^\ell+\dots+\alpha_0$ with $\theta$ being the free variable is a Hurwitz polynomial. Using the adaptive controller given by (\ref{eq:5}) and (\ref{eq:7}) yields a differential-cascaded system with degree $\ell$
\be
\label{eq:17}
\begin{cases}
\frac{d^{\ell+1}\Delta q}{dt^{\ell+1}}=-\alpha_\ell \frac{d^\ell \Delta q}{dt^\ell}-\dots-\alpha_0\Delta q\\
\qquad+\lambda_{\mathcal S}^\ast \lambda_c\hat M(q) s+\lambda_{\mathcal S}^\ast\frac{d}{dt}[\hat M(q)s]+ \frac{d^\ell s}{dt^\ell}\\
\frac{d}{dt}[M(q)s-W\Delta \vartheta]=-\lambda_c[M(q)s-W\Delta \vartheta]+\tau^\ast\\
\dot {\hat\vartheta}=-\Gamma W^T s.
\end{cases}
\ee

\emph{Theorem 3:} The adaptive controller given as (\ref{eq:5}) and (\ref{eq:7}) with $z$ given as (\ref{eq:16}) ensures the convergence of the tracking errors, i.e., $\Delta q\to 0$ and $\Delta \dot q\to 0$ as $t\to\infty$ with $\tau^\ast=0$. The quantitative nonlinearity measure of the dynamics of the transmission from $\tau^\ast$ to $\Delta q$ (with respect to linear dynamics) in the sense of certainty equivalence can be specified by the residual error ${d^\ell s}/{dt^\ell}$, $\ell=2,3,\dots$.

\emph{Proof:} Most of the proof follows similar procedures as in that of Theorem 1, and the distinctive point mainly lies in the analysis of the first subsystem of (\ref{eq:17}) with $s\in\mathcal L_2$ and $\hat\vartheta\in\mathcal L_\infty$.
The first subsystem of (\ref{eq:17}) can further be written as
\begin{align}
\label{eq:aa1}
\frac{d^{\ell} \Delta z}{dt^{\ell}}=&-\alpha_\ell \frac{d^{\ell-1} \Delta z}{dt^{\ell-1}}-\dots-\alpha_1\Delta z-\alpha_0\Delta q\nn\\
&-\alpha_\ell\frac{d^{\ell-1}s}{dt^{\ell-1}}-\dots-\alpha_1 s+\lambda_{\mathcal S}^\ast \lambda_c\hat M(q) s\nn\\
&+\lambda_{\mathcal S}^\ast\frac{d}{dt}[\hat M(q)s]
\end{align}
where $\Delta z=z-\dot q_d$. The state-space representation of (\ref{eq:aa1}) can be formulated as
\begin{align}
\label{eq:aa2}
\begin{cases}
\Delta \dot q=\Delta z+s\\
\quad\vdots\\
\frac{d}{dt}({\frac{d^{\ell-2} \Delta z}{dt^{\ell-2}}})={\frac{d^{\ell-1} \Delta z}{dt^{\ell-1}}}\\
\frac{d}{dt}(\frac{d^{\ell-1} \Delta z}{dt^{\ell-1}})=-\alpha_\ell \frac{d^{\ell-1} \Delta z}{dt^{\ell-1}}-\dots-\alpha_1\Delta z-\alpha_0\Delta q\\
\qquad-\alpha_\ell\frac{d^{\ell-1}s}{dt^{\ell-1}}-\dots-\alpha_1 s+\lambda_{\mathcal S}^\ast \lambda_c\hat M(q) s\\
\qquad+\lambda_{\mathcal S}^\ast\frac{d}{dt}[\hat M(q)s]
\end{cases}
\end{align}
with $x=[\Delta q^T,\Delta z^T,\dots,(d^{\ell-1}\Delta z/dt^{\ell-1})^T]^T$ as the state and with $u_1=[s^T,0_{n(\ell-1)}^T,[-\alpha_1 s+\lambda_{\mathcal S}^\ast\lambda_c\hat M(q)s]^T]^T$, $u_2=-\alpha_2\dot s+\lambda_{\mathcal S}^\ast\frac{d}{dt}[\hat M(q)s]$, and $u_k=-\alpha_kd^{k-1}s/dt^{k-1}$, $k=3,\dots,\ell$ as the inputs. The system (\ref{eq:aa2}) with $u_k=0$, $k=1,\dots,\ell$ is exponentially stable due to the fact that $\theta^{\ell+1}+\alpha_\ell \theta^\ell+\dots+\alpha_0$ is a Hurwitz polynomial. Using the standard input-output properties of exponentially stable and strictly proper linear systems \cite[p.~59]{Desoer1975_Book}, the part of $x$ due to the input $u_1$ is square-integrable and bounded and converges to zero since $u_1\in\mathcal L_2$. Let us now consider the mappings from $u_k$ to $\Delta q$, $\Delta z$, and $\Delta \dot z$, respectively, $k=2,\dots,\ell$, and in accordance with the standard linear system theory, the relative degree of $\Delta q$, $\Delta z$, and $\Delta \dot z$ with respect to $-\alpha_2 s+\lambda_{\mathcal S}^\ast \hat M(q)s$, $-\alpha_{\kappa} s$, $\kappa=3,\dots,\ell$ is $\ell-k+2$, $\ell-k+1$, and $\ell-k$, respectively, $k=2,\dots,\ell$. From the input-output properties of exponentially stable linear systems \cite[p.~59]{Desoer1975_Book}, we have that the parts of $\Delta q$ and $\Delta z$ due to $u_k$ are square-integrable and bounded and converges to zero, and that the part of $\Delta \dot z$ due to $u_k$ is square-integrable, $k=2,\dots,\ell$. The application of the standard superposition principle for linear systems yields the result that $\Delta q\in\mathcal L_2\cap\mathcal L_\infty$, $\Delta z\in\mathcal L_2\cap\mathcal L_\infty$, $\Delta \dot z\in\mathcal L_2$, and $\Delta q\to 0$ as $t\to\infty$. Hence, $\Delta \dot q=\Delta z+s\in\mathcal L_2$. The remaining proof can be completed with similar procedures as in the proof of Theorem 1. \hfill \text{\small $\blacksquare$}

\emph{Remark 2:} The system (\ref{eq:17}) exhibits properties that have certain similarity to the standard approximation of functions using Taylor's theorem in calculus. For a nonlinear and uncertain Lagrangian system such as a robot manipulator, using forwardstepping design \cite{Wang2019_ACC,Wang2020_arXiv} concerning the reference dynamics and the inertia-invariance-based torque input design concerning the handling of nonlinearity and parametric uncertainty leads us to render the closed-loop dynamics approximately behave like linear dynamics with a high-order residual or approximation error [for instance, the term $d^\ell s/dt^\ell$ in (\ref{eq:17})]. Since the late 1980s, the Slotine and Li adaptive controller \cite{Slotine1987_IJRR} has become a standard for the control of robot manipulators and general Lagrangian systems, and this controller has been argued as ``putting physics in control'' (see \cite{Slotine1988_CSM}). In other words, since the Slotine and Li controller, the possible connection between control of robot manipulators and the linear control has almost been abandoned (due also in part to the limitations of the adaptive inverse dynamics or feedback linearization \cite{Craig1987_IJRR,Middleton1988_SCL}, either involving inversion/invertibility of the estimated inertia or joint acceleration measurement), and the occurrence of passivity-based control also happens in this context (see \cite{Ortega1989_AUT}). The presented forwardstepping design here establishes this connection that has long been considered as intractable, which may also have the potential of becoming a new systematic approach for control of nonlinear systems. In contrast with some standard nonlinear design approaches such as backstepping \cite{Krstic1995_Book}, forwarding \cite{Teel1992_IFAC,Sepulchre1996_IFAC}, or the immersion and invariance (I\&I) approach \cite{Astolfi2003_TAC}, typically yielding cascaded closed-loop dynamics (with reduced-order reference dynamics), the presented design methodology approaches the nonlinear control problem along a fundamentally different direction using reference dynamics with an increment of the order.

\emph{Remark 3:} In the traditional adaptive inverse dynamics or feedback linearization of robot manipulators, typically a regressor matrix that differs from the above formulation is involved \cite{Craig1987_IJRR,Middleton1988_SCL,Li1989_SCL}, namely the regressor matrix $Y^\ast(q,\dot q,\ddot q)$ defined by
\be
M(q)\ddot q+C(q,\dot q)\dot q+g(q)=Y^\ast(q,\dot q,\ddot q)\vartheta.
\ee
We reformulate these inverse dynamics controllers in the context of forwardstepping design that is compatible with degree reduction analysis. Upon the regressor matrix $Y^\ast$, we define a matrix $W^{\ast}$ using the filtering technique as \cite{Middleton1988_SCL,Slotine1991_Book}
\be
\dot W^\ast=-\lambda_c W^\ast+Y^\ast(q,\dot q,\ddot q)
\ee
and the matrix $W^\ast$ can be calculated without involving joint acceleration measurement in accordance with \cite{Middleton1988_SCL,Slotine1991_Book}.
The torque control given by (\ref{eq:5}) needs to be modified as (similar to the torque input given in \cite{Li1989_SCL})
\begin{align}
\label{eq:aa3}
\tau=&-\lambda_c \hat M(q) s+\hat M(q)\dot z+\hat C(q,\dot q)\dot q\nn\\
&+\hat g(q)-\dot {\hat M}(q)s+W^\ast\dot{\hat \vartheta}
\end{align}
and this yields dynamics that differ from the second subsystem of (\ref{eq:9}) as (which is obtained by degree reduction concerning $\frac{d}{dt}[W^\ast\Delta\vartheta]$)
\be
\frac{d}{dt}[\hat M(q)s-W^\ast\Delta \vartheta]=-\lambda_c [\hat M(q)s- W^\ast\Delta\vartheta]
\ee
which involves the estimated inertia matrix. At this stage, we can develop the parameter adaptation law for $\hat\vartheta$ along two directions, as is typically done, namely direct and indirect adaptation, respectively. The direct adaptation law can be given as
\be
\label{eq:aa4}
\dot{\hat\vartheta}=-\Gamma W^{\ast T}s
\ee
or as
\be
\dot{\hat\vartheta}=-\Gamma W^{\ast T}\hat M(q)s.
\ee
The indirect adaptation law can be specified to be the same as \cite{Middleton1988_SCL,Li1989_SCL}, i.e.,
\be
\dot{\hat\vartheta}=-\Gamma W^{\ast T}e
\ee
with $e=W^\ast\hat\vartheta-\tau_f$ being the prediction error and $\tau_f$ given as $\dot \tau_f=-\lambda_c\tau_f+\tau$. The reformulated direct adaptive inverse dynamics approach does not rely on the joint acceleration measurement, in contrast with \cite{Craig1987_IJRR}. The stability of both the direct and indirect approaches, due to the involvement of the estimated inertia matrix, relies on the condition that the estimated inertia $\hat M(q)$ is invertible (or uniformly positive definite), similar to the standard direct and indirect inverse dynamics approaches in \cite{Craig1987_IJRR,Middleton1988_SCL,Li1989_SCL}. The invertibility of $\hat M(q)$ is generally considered to be most intractable in the traditional inverse dynamics framework with either a direct or indirect approach (see, e.g., \cite{Spong1990_TAC}). The connection presented here indicates how the forwardstepping approach with inertia invariance differs from the standard adaptive inverse dynamics approach, and why the inversion/invertibility of the estimated inertia and joint acceleration measurement are no longer required in the proposed adaptive approach (due to inertia invariance and degree-reduction-based formulation, respectively). In fact, with the implementation of the proposed inertia invariance and degree-reduction-based formulation, the adaptive inverse dynamics control given by (\ref{eq:aa3}) and (\ref{eq:aa4}) can be transformed to the proposed control given by (\ref{eq:5}) and (\ref{eq:7}). Without considering the introduction of reference dynamics based upon the order invariance or increment, the presented control has some similarity to \cite{Seo2009_SCL} yet exhibits a much compact structure due to the analysis based on degree reduction and a quasi-Lyapunov function; in addition, the presented analysis/design, as demonstrated above, facilitates establishing the connection with the traditional adaptive inverse dynamics approaches.

\section{Adaptive Control with Constant-Gain Feedback}

Constant-gain feedback is preferred for its enhanced robustness in comparison with the time-varying-gain feedback relying on the estimated inertia. This is mainly due to the well-recognized robustness issue associated with adaptive control in the context of external disturbances or input.

\subsection{Constant-Gain Feedback}

For this purpose, we define reference dynamics as
\begin{align}
\label{eq:25}
\frac{d^\ell z}{dt^\ell}=&\frac{d^{\ell+1}q_d}{dt^{\ell+1}}-\alpha_\ell \frac{d^\ell \Delta q}{dt^\ell}-\dots-\alpha_0\Delta q\nn\\
&+\lambda_{\mathcal S}^\ast\frac{d}{dt}[\hat M(q) (\dot q-z)]+\lambda_{\mathcal S}^\ast \lambda_c^\ast(\dot q-z)
\end{align}
where $\lambda_c^\ast$ is a positive design constant. The control torque is given as
\begin{align}
\label{eq:26}
\tau=&-\lambda_c^\ast s+\hat M(q)\dot z+\hat C(q,\dot q)\dot q\nn\\
&+\hat g(q)-\hat {\dot M}(q)s+W\dot{\hat \vartheta}
\end{align}
where $W$ is given by (\ref{eq:6}) with
 the regressor $Y(q,\dot q,z,\dot z)$ being redefined as [differing from (\ref{eq:4})]
\be
\label{eq:27}
 M(q)\dot z+ C(q,\dot q)\dot q+ g(q)- {\dot M}(q)s=Y(q,\dot q,z,\dot z)\vartheta.
\ee

The closed-loop dynamics of the manipulator can be described by a differential-cascaded system as
\be
\label{eq:28}
\begin{cases}
\frac{d^{\ell+1}\Delta q}{dt^{\ell+1}}=-\alpha_\ell \frac{d^\ell \Delta q}{dt^\ell}-\dots-\alpha_0\Delta q\\
\qquad+\lambda_{\mathcal S}^\ast \lambda_c^\ast s+\lambda_{\mathcal S}^\ast\frac{d}{dt}[\hat M(q)s]+ \frac{d^\ell s}{dt^\ell}\\
\frac{d}{dt}[M(q)s-W\Delta \vartheta]=-\lambda_c^\ast s+\lambda_c W\Delta \vartheta+\tau^\ast\\
\dot {\hat\vartheta}=-\Gamma W^T s.
\end{cases}
\ee
Consider the Lyapunov-like function candidate
\begin{align}
V^\ast=&(1/2)[M(q)s-W\Delta\vartheta]^T[M(q)s-W\Delta\vartheta]\nn\\
&+(\lambda_c^\ast/2)\Delta\vartheta^T\Gamma^{-1}\Delta\vartheta
\end{align}
and its derivative along the trajectories of the system can be written as
\begin{align}
\label{eq:30}
\dot V^\ast=&-\lambda_c^\ast s^T M(q)s+\lambda_c s^T M(q)W\Delta \vartheta\nn\\
&-\lambda_c\Delta\vartheta^T W^T W\Delta\vartheta.
\end{align}
To facilitate the specification of the controller parameters, we rewrite (\ref{eq:30}) as
\begin{align}
\dot V^\ast=&-s^TM(q)[\lambda_c^\ast I_m-(1/4)\lambda_c M(q)]s\nn\\
&-\lambda_c[(1/2)M(q)s-W\Delta\vartheta]^T \nn\\
&\times[(1/2)M(q)s-W\Delta\vartheta]\le 0
\end{align}
if $\lambda_c^\ast>(1/4)\lambda_c\lambda_{\max}\{M(q)\}$, where $\lambda_{\max}\{\cdot\}$ denotes the maximum eigenvalue of a matrix.

\emph{Theorem 4:} Let the controller parameters $\lambda_c^\ast$ and $\lambda_c$ be chosen such that
$
\lambda_c^\ast>(1/4)\lambda_c\lambda_{\max}\{M(q)\}
$. Then, the adaptive controller given as (\ref{eq:26}) and (\ref{eq:7}) with $z$ given as (\ref{eq:25}) and $Y(q,\dot q,z,\dot z)$ given as (\ref{eq:27}) ensures the convergence of the tracking errors, i.e., $\Delta q\to 0$ and $\Delta \dot q\to 0$ as $t\to\infty$ with $\tau^\ast=0$. The quantitative nonlinearity measure of the dynamics of the transmission from $\tau^\ast$ to $\Delta q$ (with respect to linear dynamics) in the sense of certainty equivalence can be specified by the residual error ${d^\ell s}/{dt^\ell}$, $\ell=2,3,\dots$.

\subsection{Constant-Gain Feedback without Relying on the Knowledge of $M(q)$}

The control torque is specified as
\begin{align}
\label{eq:32}
\tau=&-\lambda_c^\ast s+\hat M(q)\dot z+\hat C(q,\dot q)\dot q\nn\\
&+\hat g(q)-\dot {\hat M}(q)s+W^\ast\dot{\hat \vartheta}
\end{align}
and the adaptation law is given as
\be
\label{eq:33}
\dot{\hat\vartheta}=-\Gamma[\lambda_cW^{\ast T}\hat M(q)s+\lambda_c^\ast W^{\ast\ast T}s]
\ee
where $W^{\ast\ast}$ is defined as
\be
W^{\ast\ast}\vartheta=W^\ast \vartheta-M(q)s.
\ee
The closed-loop dynamics of the manipulator can be described by a differential-cascaded system as
\be
\label{eq:35}
\begin{cases}
\frac{d^{\ell+1}\Delta q}{dt^{\ell+1}}=-\alpha_\ell \frac{d^\ell \Delta q}{dt^\ell}-\dots-\alpha_0\Delta q\\
\qquad+\lambda_{\mathcal S}^\ast \lambda_c^\ast s+\lambda_{\mathcal S}^\ast\frac{d}{dt}[\hat M(q)s]+ \frac{d^\ell s}{dt^\ell}\\
\frac{d}{dt}[\hat M(q)s-W^\ast\Delta \vartheta]=-\lambda_c^\ast s+\lambda_c W^\ast\Delta \vartheta+\tau^\ast\\
\dot{\hat\vartheta}=-\Gamma[\lambda_cW^{\ast T}\hat M(q)s+\lambda_c^\ast W^{\ast\ast T}s].
\end{cases}
\ee

\emph{Theorem 5:} The adaptive controller given as (\ref{eq:32}) and (\ref{eq:33}) with $z$ given as (\ref{eq:25}) ensures the convergence of the tracking errors, i.e., $\Delta q\to 0$ and $\Delta \dot q\to 0$ as $t\to\infty$ with $\tau^\ast=0$. The quantitative nonlinearity measure of the dynamics of the transmission from $\tau^\ast$ to $\Delta q$ (with respect to linear dynamics) in the sense of certainty equivalence can be specified by the residual error ${d^\ell s}/{dt^\ell}$, $\ell=2,3,\dots$.

The proof of Theorem 5 can be completed using the Lyapunov-like function candidate $
V^\ast=(1/2)[\hat M(q)s-W^\ast\Delta\vartheta]^T[\hat M(q)s-W^\ast\Delta\vartheta]+(1/2)\Delta\vartheta^T\Gamma^{-1}\Delta \vartheta
$ and similar procedures as previously conducted.

\emph{Remark 4:} In many interactive control problems such as teleoperation or teaching operation, the reference input torque $\tau^\ast$ typically covers the low-frequency range. To ensure the reliability and robustness of the system under the action of $\tau^\ast$, it tends to be preferable that the adaptation becomes weaker or even vanishes with $\tau^\ast$ being static (typically, the joint velocity $\dot q$ becomes zero). This is achieved in the system given by (\ref{eq:28}) yet not in the system given by (\ref{eq:35}), and the price is that the controller parameters rely on some a priori knowledge of $M(q)$.

\section{Choice and Implementation of Reference Dynamics}

\subsection{Choice of Reference Dynamics}

The order invariance and increment of the proposed approach allows much freedom for the choice of reference dynamics. Though there are indeed many reference dynamics that can guarantee the stability and convergence of the closed-loop systems, different reference dynamics might lead to systems with strong or weak robustness and different performance. We illustrate this issue via the example given in Sec. 3.1. Consider the system (\ref{eq:9}), and it can be rewritten as (with degree reduction concerning the first subsystem)
\be
\begin{cases}
\frac{d}{dt}(\Delta\dot q-s)=-\alpha_1\Delta \dot q-\alpha_0\Delta q+\lambda_{\mathcal S}^\ast \lambda_c\hat M(q) s\\
\frac{d}{dt}[M(q)s-W\Delta \vartheta]=-\lambda_c[M(q)s-W\Delta \vartheta]+\tau^\ast\\
\dot {\hat\vartheta}=-\Gamma W^T s.
\end{cases}
\ee
From an input-output perspective, the first subsystem is not so ``symmetric'' in the sense that with $\lambda_{\mathcal S}^\ast\lambda_c\hat M(q)s=0$, it is not a typical linear system due to the involvement of $s$ in $\frac{d}{dt}(\Delta \dot q-s)$. Let us now redefine the reference dynamics as
\begin{align}
\dot z=&\ddot q_d-\alpha\Delta \dot q-\Lambda(z-\dot q_d)-\alpha\Lambda\Delta q\nn\\
&+\lambda_{\mathcal S}^\ast\lambda_c \hat M(q)(\dot q-z)
\end{align}
with $\alpha$ being a positive design constant and $\Lambda$ a symmetric positive definite matrix, and we have a differential-cascaded system differing from (\ref{eq:9})
\be
\begin{cases}
\frac{d}{dt}(\Delta\dot q+\alpha\Delta q-s)=-\Lambda(\Delta\dot q+\alpha\Delta q-s)+\lambda_{\mathcal S}^\ast \lambda_c\hat M(q) s\\
\frac{d}{dt}[M(q)s-W\Delta \vartheta]=-\lambda_c[M(q)s-W\Delta \vartheta]+\tau^\ast\\
\dot {\hat\vartheta}=-\Gamma W^T s.
\end{cases}
\ee
This differential-cascaded system is, in some sense, ``symmetric'' since in the case that $\lambda_{\mathcal S}^\ast \lambda_c\hat M(q) s=0$, the first subsystem is indeed a linear system with respect to the state $\Delta \dot q+\alpha\Delta q-s$. The other examples in Sec. 3 can be reshaped with a similar procedure.

\subsection{Implementation of Reference Dynamics}

Many reference dynamics, from their definitions, involve challenging measurement or calculation of functions or variables, e.g., the joint acceleration, its derivative, or its high-order derivatives and the derivative of the estimated inertia matrix $\hat M(q)$. On the other hand, we note that only $z$ and $\dot z$ are used in the proposed adaptive controllers (for a second-order robot manipulator). Here, we demonstrate in detail how the joint acceleration and its derivative or high-order derivatives are avoided as calculating $z$ and $\dot z$ numerically. We first consider the reference dynamics given by (\ref{eq:12}), and this system can be considered as a linear time-varying system concerning $z$, $\dot z$, and $\ddot z$. Hence, we cannot directly employ the filtering technique in \cite{Slotine1991_Book} for the linear time-invariant case. For handling this issue, we apply the degree reduction \cite{Wang2019_ACC} (in particular, concerning the derivative or high-order derivatives of $\dot q$) to the linear time-varying system (\ref{eq:12}) and have that
 \begin{align}
\label{eq:aa5}
\begin{cases}
\dot z=-\lambda_{\mathcal S}^\ast\hat M(q)z+[\dot z-\ddot q_d+\lambda_{\mathcal S}^\ast\hat M(q)(z-\dot q)+\alpha_2 \dot q]\\
\qquad+\lambda_{\mathcal S}^\ast\hat M(q)\dot q+\ddot q_d-\alpha_2\dot q\\
\frac{d}{dt}[\dot z-\ddot q_d+\lambda_{\mathcal S}^\ast\hat M(q)(z-\dot q)+\alpha_2\dot q]\\
=-\lambda_c\lambda_{\mathcal S}^\ast \hat M(q)z+\alpha_2\ddot q_d-\alpha_1\Delta \dot q-\alpha_0\Delta q\\
\quad+\lambda_c\lambda_{\mathcal S}^\ast \hat M(q)\dot q
\end{cases}
\end{align}
whose state-space form with $x^\ast=[z^T,[\dot z-\ddot q_d+\lambda_{\mathcal S}^\ast\hat M(q)(z-\dot q)+\alpha_2\dot q]^T]^T$ can be written as
\begin{align}
\label{eq:aa6}
\dot x^\ast=A x^\ast+u^\ast
\end{align}
where $u^\ast=[u_1^{\ast T},u_2^{\ast T}]^T$ with $u_1^\ast=\lambda_{\mathcal S}^\ast\hat M(q)\dot q+\ddot q_d-\alpha_2\dot q$ and $ u_2^\ast=\alpha_2 \ddot q_d-\alpha_1\Delta \dot q-\alpha_0\Delta q+\lambda_c\lambda_{\mathcal S}^\ast \hat M(q)\dot q$, and
$A=\begin{bmatrix}-\lambda_{\mathcal S}^\ast \hat M(q)& I_n\\
-\lambda_c\lambda_{\mathcal S}^\ast \hat M(q) & 0_{n\times n}\end{bmatrix}$. By the numerical integral operation of (\ref{eq:aa6}), which no longer involves the joint acceleration measurement and $\dddot q_d$, we can obtain $x^\ast$, and then $z$ and $\dot z$.
Let us further consider the reference dynamics (\ref{eq:16}) with $\ell=3$
\begin{align}
\label{eq:aa7}
\dddot z=&\ddddot q_d-\alpha_3\Delta \dddot q-\alpha_2\Delta \ddot q-\alpha_1\Delta \dot q-\alpha_0\Delta q\nn\\
&+\lambda_{\mathcal S}^\ast\frac{d}{dt}[\hat M(q)\dot q]+\lambda_c\lambda_{\mathcal S}^\ast \hat M(q)\dot q\nn\\
&-\lambda_{\mathcal S}^\ast\frac{d}{dt}[\hat M(q) z]-\lambda_c\lambda_{\mathcal S}^\ast \hat M(q) z.
\end{align}
Using degree reduction to (\ref{eq:aa7}) yields
\begin{align}
\label{eq:aa8}
\begin{cases}
\dot z=(\dot z-\ddot q_d+\alpha_3\dot q)+\ddot q_d-\alpha_3\dot q\\
\frac{d}{dt}(\dot z-\ddot q_d+\alpha_3\dot q)\\
=[\ddot z-\dddot q_d+\lambda_{\mathcal S}^\ast\hat M(q)(z-\dot q)+\alpha_3\Delta\ddot q+\alpha_2\dot q]\\
\qquad-\lambda_{\mathcal S}^\ast \hat M(q)z+\lambda_{\mathcal S}^\ast \hat M(q)\dot q+\alpha_3\ddot q_d-\alpha_2\dot q\\
\frac{d}{dt}[\ddot z-\dddot q_d+\lambda_{\mathcal S}^\ast\hat M(q)(z-\dot q)+\alpha_3\Delta\ddot q+\alpha_2\dot q]\\
=-\lambda_c\lambda_{\mathcal S}^\ast \hat M(q) z+\alpha_2 \ddot q_d-\alpha_1\Delta \dot q-\alpha_0\Delta q\\
\quad+\lambda_c\lambda_{\mathcal S}^\ast \hat M(q)\dot q.
\end{cases}
\end{align}
The state for this linear time-varying system can be chosen as $x^\ast=[z^T,(\dot z-\ddot q_d+\alpha_3\dot q)^T,[\ddot z-\dddot q_d+\lambda_{\mathcal S}^\ast\hat M(q)(z-\dot q)+\alpha_3\Delta\ddot q+\alpha_2\dot q]^T]^T$. The numerical integration of (\ref{eq:aa8}) relies on the specification of the initial value of $\ddot q$, namely $\ddot q(0)$, and its value is difficult to be known exactly; however, this value can be chosen to be an arbitrary constant since this value does not influence the holding of (\ref{eq:aa7}). It is similar for the involvement of $\dddot q_d$, and neither this quantity nor its initial value is required to be known. On the other hand, the involvement of such quantities in $x^\ast(0)$ can also be avoided via a particular choice of $\dot z(0)$ and $\ddot z(0)$ and exploiting the fact that only $x^\ast(0)$ is required for implementation. For instance, consider the reference dynamics given as (\ref{eq:aa7}) with $\alpha_0=\alpha^4$, $\alpha_1=4\alpha^3$, $\alpha_2=6\alpha^2$, and $\alpha_3=4\alpha$ where $\alpha$ is a positive design constant, and we choose $\ddot z(0)=\dddot q_d(0)-3\alpha\Delta \ddot q(0)-3\alpha^2\Delta \dot q(0)-\alpha^3\Delta q(0)$; then, $x^\ast_3(0)=[\ddot z-\dddot q_d+\lambda_{\mathcal S}^\ast\hat M(q)(z-\dot q)+\alpha_3\Delta\ddot q+\alpha_2\dot q]|_{t=0}=\alpha\Delta \ddot q(0)+6\alpha^2\dot q(0)-3\alpha^2\Delta \dot q(0)-\alpha^3\Delta q(0)+\lambda_{\mathcal S}^\ast\hat M(q(0))[z(0)-\dot q(0)]$, which no longer involves the information concerning $\dddot q_d$. This indicates that the integral operation of the degree-reduction form of the reference dynamics does not rely on the high-order information such as $\dddot q_d$. The arbitrary-order reference dynamics can similarly be accommodated using degree reduction (which would certainly become more complicated as the order increases).

\emph{Remark 5:} The degree-reduction-based implementation of reference dynamics given here renders it practically possible to achieve the objective that the controlled nonlinear robots approach linear dynamics using position and velocity measurement only. This significantly facilitates the practical and reliable applications of the proposed controllers to robot manipulators with quantifiable performance. The comparison between the degree-reduction-based schemes and their original definitions indicates that the integral stability and accuracy is maintained without relying on challenging measurement of variables (e.g., joint acceleration, its derivative or high-order derivatives) due to the degree-reduction-based realization.

\section{Simulation Results}

We consider the application of the proposed adaptive controllers to a standard two-DOF (degree-of-freedom) planar manipulator motioning in a horizontal plane. The dynamics of the manipulator can be found in \cite{Spong2006_Book}. The physical parameters of the manipulator (assume that the mass density of the links is uniform along the length direction) are set as $m_1=3.6$, $m_2=2.7$, $l_1=1.8$, and $l_2=1.8$ where $m_i$ and $l_i$ are the mass and length of the $i$-th link, respectively, $i=1,2$. The sampling period is specified as 5 ms.

We set the controller parameters as $\lambda_c=10$, $\lambda_{\mathcal S}^\ast=0.5$, and $\Gamma=10I_3$. The initial parameter estimate is set as $\hat\vartheta(0)=[0,0,0]^T$. For the controller with differential-cascaded-degree one, the controller parameters are chosen as $\alpha_0=100$ and $\alpha_1=2\sqrt{\alpha_0}$ while $z(0)$ is set as $z(0)=\dot q_d(0)$; for the controller with differential-cascaded-degree two, we choose $\alpha_0=100$, $\alpha_1=3\alpha_0^{2/3}$, and $\alpha_2=3\alpha_0^{1/3}$, and the initial values for $z$ and $\dot z$ are set as $z(0)=\dot q_d(0)$ and $\dot z(0)=\ddot q_d(0)-2\alpha\Delta \dot q(0)-\alpha^2\Delta q(0)$, respectively with $\alpha=\alpha_0^{1/3}$; for the controller with differential-cascaded-degree three, we choose $\alpha_0=100$, $\alpha_1=4\alpha_0^{3/4}$, $\alpha_2=6\alpha_0^{1/2}$, and $\alpha_3=4\alpha_0^{1/4}$, and the initial values for $z$, $\dot z$, and $x_3^\ast$ are set as $z(0)=\dot q_d(0)$, $\dot z(0)=\ddot q_d(0)-3\alpha\Delta \dot q(0)-3\alpha^2\Delta q(0)$, and $x_3^\ast(0)=\alpha[\dot z(0)-\ddot q_d(0)]+6\alpha^2\dot q(0)-3\alpha^2\Delta \dot q(0)-\alpha^3\Delta q(0)+\lambda_{\mathcal S}^\ast\hat M(q(0))[z(0)-\dot q(0)]$, respectively with $\alpha=\alpha_0^{1/4}$. With such choice of controller parameters, the three controllers maintain the same low-frequency gain from the reference torque $\tau^\ast$ to the position tracking error $\Delta q$, namely $\lambda_{\mathcal S}^\ast/\alpha_0$, and their difference lies in the nonlinearity measure of the closed-loop dynamics quantified by a remainder. The desired joint position is specified as $q_d=[(\pi/3)\sin(\pi t),(\pi/3)\sin(\pi t)]^T$. The tracking errors with the first controller being implemented are shown in Fig. 1 where the low-frequency part of the controlled robot dynamics $\lambda_{\mathcal S}^\ast\lambda_c\hat M(q)s$ is incorporated in the reference dynamics (without considering the remainder $\dot s$). The tracking errors with the second and third controllers (whose reference dynamics incorporate the full part of the controlled robot dynamics, i.e., $\lambda_{\mathcal S}^\ast\lambda_c\hat M(q)s+\lambda_{\mathcal S}^\ast\frac{d}{dt}[\hat M(q)s]$, differing from the first controller) being implemented are shown in Fig. 2 and Fig. 3, respectively. It can be observed from Fig. 2 and Fig. 3 that with the increment of the differential-cascaded degree (from degree 2 to degree 3), the response of the tracking errors is improved in both the smoothness and tracking accuracy.

We also conduct the simulation study with the reference dynamics being modified using similar procedures as in Sec. 5.1. The modified reference dynamics for the adaptive controllers with differential-cascaded degree two and three can be respectively written as
 \begin{align}
\ddot z=&\dddot q_d-2\alpha\Delta \ddot q-\alpha^2\Delta \dot q-\Lambda(\dot z-\ddot q_d)-2\alpha\Lambda\Delta \dot q-\alpha^2\Lambda\Delta q\nn\\
&+\lambda_{\mathcal S}^\ast\frac{d}{dt}[\hat M(q)(\dot q-z)]+\lambda_{\mathcal S}^\ast\lambda_c \hat M(q)(\dot q-z)
\end{align}
and
 \begin{align}
\dddot z=&\ddddot q_d-3\alpha\Delta \dddot q-3\alpha^2\Delta\ddot q-\alpha^3\Delta \dot q\nn\\
&-\Lambda(\ddot z-\dddot q_d)-3\alpha\Lambda\Delta \ddot q-3\alpha^2\Lambda\Delta \dot q-\alpha^3\Lambda \Delta q\nn\\
&+\lambda_{\mathcal S}^\ast\frac{d}{dt}[\hat M(q)(\dot q-z)]+\lambda_{\mathcal S}^\ast\lambda_c \hat M(q)(\dot q-z).
\end{align}
These reference dynamics yield
 \begin{align}
\frac{d}{dt}&(\Delta \ddot q+2\alpha\Delta \dot q+\alpha^2\Delta q-\dot s)\nn\\
=&-\Lambda(\Delta \ddot q+2\alpha\Delta \dot q+\alpha^2\Delta q-\dot s)\nn\\
&+\lambda_{\mathcal S}^\ast\frac{d}{dt}[\hat M(q)s]+\lambda_{\mathcal S}^\ast\lambda_c \hat M(q)s
\end{align}
and
 \begin{align}
\frac{d}{dt}&(\Delta \dddot q+3\alpha\Delta \ddot q+3\alpha^2\Delta\dot q+\alpha^3\Delta q-\ddot s)\nn\\
=&-\Lambda(\Delta \dddot q+3\alpha\Delta \ddot q+3\alpha^2\Delta\dot q+\alpha^3\Delta q-\ddot s)\nn\\
&+\lambda_{\mathcal S}^\ast\frac{d}{dt}[\hat M(q)s]+\lambda_{\mathcal S}^\ast\lambda_c \hat M(q)s.
\end{align}
The matrix $\Lambda$ is set as $\Lambda=\alpha I_2$ and the other controller parameters are chosen to be the same as the previous case.
 The simulation results with differential-cascaded degree two and three are shown in Fig. 4 and Fig. 5, respectively. In comparison with Fig. 2 and Fig. 3, the implementation of the modified reference dynamics indeed yields improved performance.

\begin{figure}
\begin{center}
\includegraphics[width=8cm]{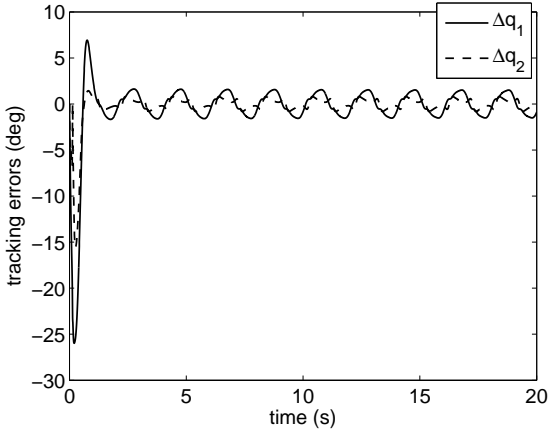}
\caption{Position tracking errors (adaptive control with $\ell=1$).}
\end{center}
\end{figure}

\begin{figure}
\begin{center}
\includegraphics[width=8cm]{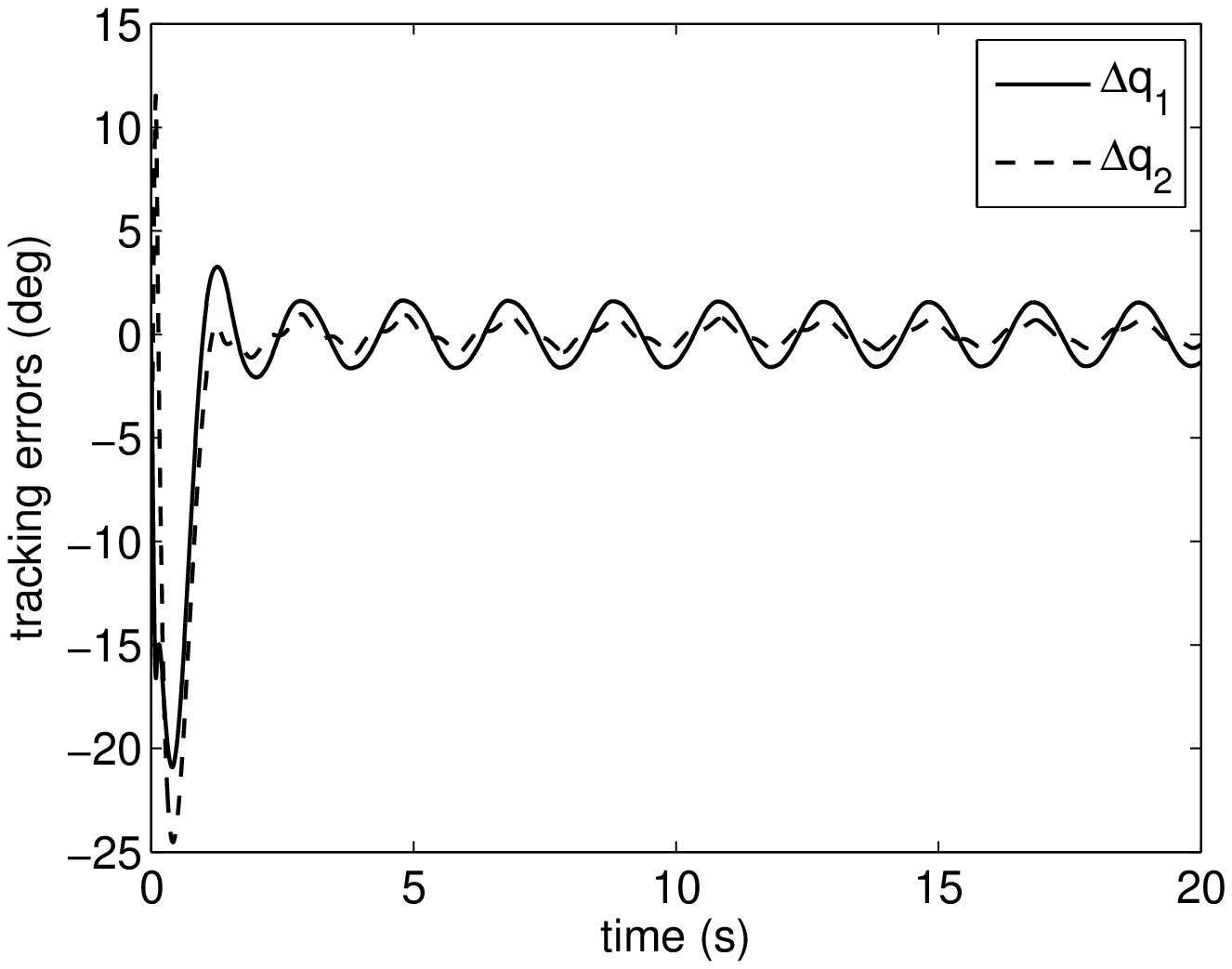}
\caption{Position tracking errors (adaptive control with $\ell=2$).}
\end{center}
\end{figure}

\begin{figure}
\begin{center}
\includegraphics[width=8cm]{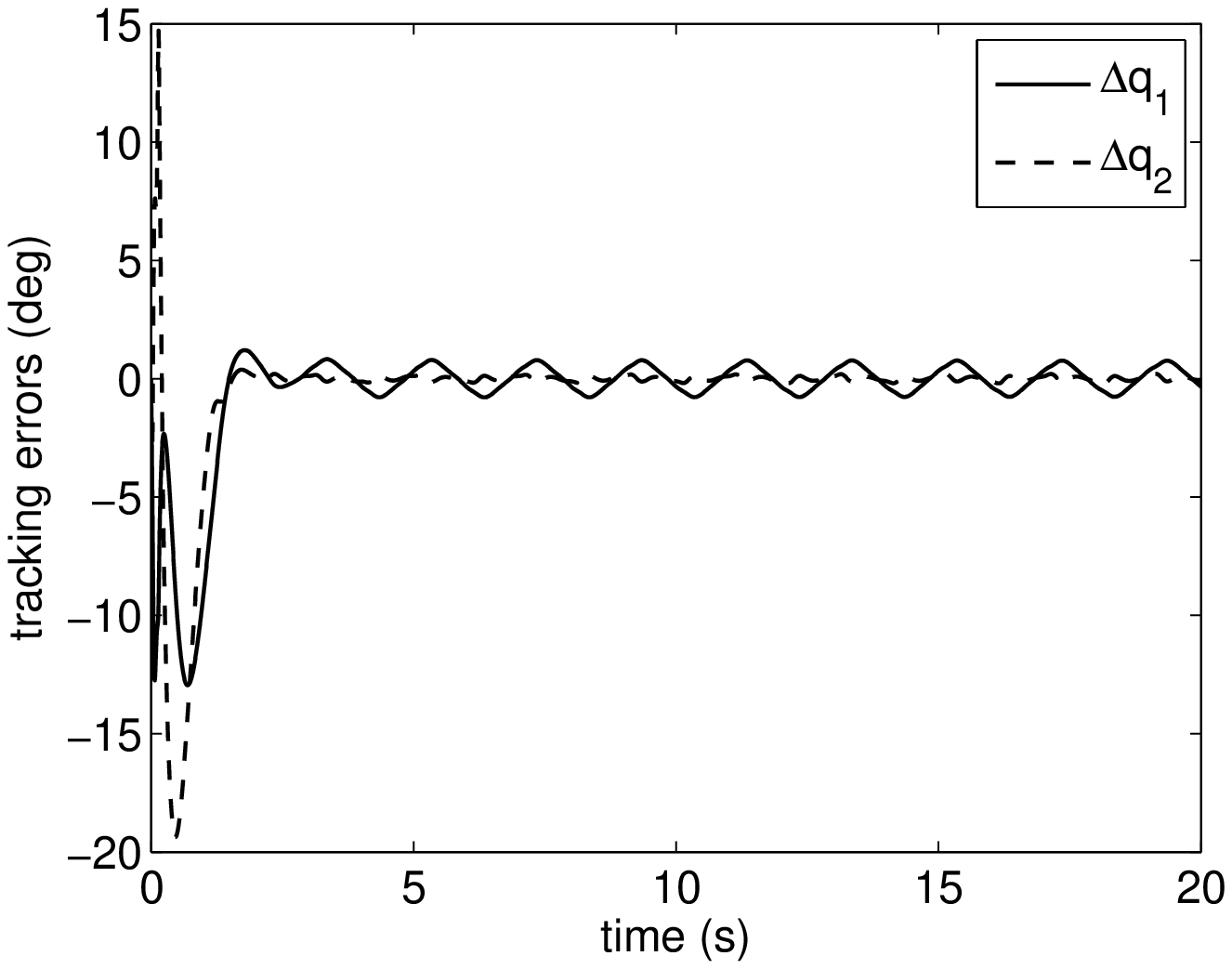}
\caption{Position tracking errors (adaptive control with $\ell=3$).}
\end{center}
\end{figure}

\begin{figure}
\begin{center}
\includegraphics[width=8cm]{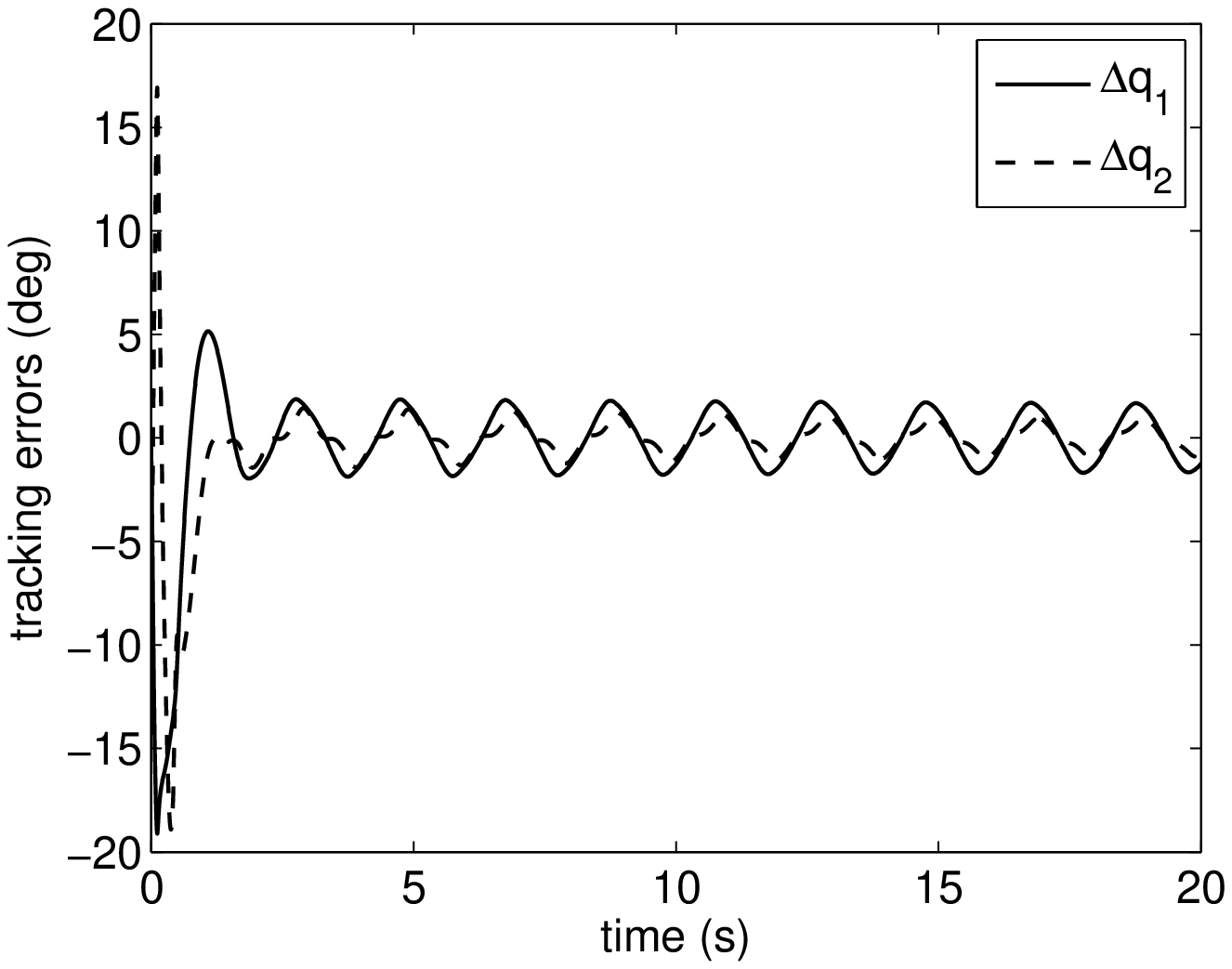}
\caption{Position tracking errors (adaptive control with $\ell=2$ and modified reference dynamics).}
\end{center}
\end{figure}

\begin{figure}
\begin{center}
\includegraphics[width=8cm]{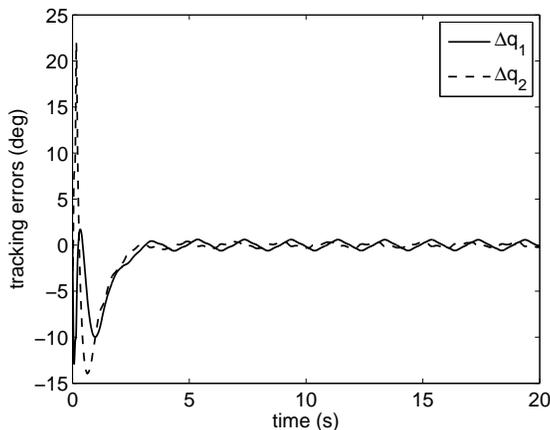}
\caption{Position tracking errors (adaptive control with $\ell=3$ and modified reference dynamics).}
\end{center}
\end{figure}

\section{Conclusion}

In this paper, we have proposed a differential-cascaded approach for the adaptive control of nonlinear robot manipulators. With the exploitation of inertia invariance and forwardstepping design of the reference dynamics, the proposed approach establishes the connection between the controlled nonlinear robot dynamics and linear dynamics, mimicking the standard Taylor's theorem in calculus for approaching functions using polynomials and their derivative and high-order derivatives. The connection between the adaptive control of nonlinear uncertain systems such as robot manipulators and linear dynamics is a long-standing intractable problem in the literature. Our result presents a solution to this problem benefiting from the exploitation of inertia invariance and the application of forwardstepping. This yields a systematic procedure to approach the nonlinear robot dynamics with linear dynamics, which also renders it possible to systematically quantify the performance of nonlinear robot dynamics with respect to a reference torque (for instance, a disturbance torque).


\bibliographystyle{plain}
\bibliography{..//Reference}

\begin{thebibliography}{10}

\bibitem{Astolfi2003_TAC}
A.~Astolfi and R.~Ortega.
\newblock Immersion and invariance: {A} new tool for stabilization and adaptive
  control of nonlinear systems.
\newblock {\em IEEE Transactions on Automatic Control}, 48(4):590--606, Apr.
  2003.

\bibitem{Berghuis1993_TRA}
H.~Berghuis and H.~Nijmeijer.
\newblock A passivity approach to controller-observer design for robots.
\newblock {\em IEEE Transactions on Robotics and Automation}, 9(6):740--754,
  Dec. 1993.

\bibitem{Cheah2003_TRA}
C.~C. Cheah, M.~Hirano, S.~Kawamura, and S.~Arimoto.
\newblock Approximate {J}acobian control for robots with uncertain kinematics
  and dynamics.
\newblock {\em IEEE Transactions on Robotics and Automation}, 19(4):692--702,
  Aug. 2003.

\bibitem{Cheah2006_IJRR}
C.~C. Cheah, C.~Liu, and J.-J.~E. Slotine.
\newblock Adaptive tracking control for robots with unknown kinematic and
  dynamic properties.
\newblock {\em The International Journal of Robotics Research}, 25(3):283--296,
  Mar. 2006.

\bibitem{Craig1987_IJRR}
J.~J. Craig, P.~Hsu, and S.~S. Sastry.
\newblock Adaptive control of mechanical manipulators.
\newblock {\em The International Journal of Robotics Research}, 6(2):16--28,
  Jun. 1987.

\bibitem{Desoer1975_Book}
C.~A. Desoer and M.~Vidyasagar.
\newblock {\em Feedback Systems: Input-Output Properties}.
\newblock Academic Press, New York, 1975.

\bibitem{Dixon2007_TAC}
W.~E. Dixon.
\newblock Adaptive regulation of amplitude limited robot manipulators with
  uncertain kinematics and dynamics.
\newblock {\em IEEE Transactions on Automatic Control}, 52(3):488--493, Mar.
  2007.

\bibitem{Egeland1994_TAC}
O.~Egeland and J.-M. Godhavn.
\newblock Passivity-based adaptive attitude control of a rigid spacecraft.
\newblock {\em IEEE Transactions on Automatic Control}, 39(4):842--846, Apr.
  1994.

\bibitem{Krstic1995_Book}
M.~Krsti{\'c}, I.~Kanellakopoulos, and P.~V. Kokotovi{\'c}.
\newblock {\em Nonlinear and Adaptive Control Design}.
\newblock Wiley, New York, 1995.

\bibitem{Li1989_SCL}
W.~Li and J.-J.~E. Slotine.
\newblock An indirect adaptive robot controller.
\newblock {\em Systems \& Control Letters}, 12(3):259--266, Apr. 1989.

\bibitem{Li2021_TCST}
Y.~Li, H.~Wang, Y.~Xie, C.~C. Cheah, and W.~Ren.
\newblock Adaptive image-space regulation for robotic systems.
\newblock {\em IEEE Transactions on Control Systems Technology},
  29(2):850--857, Mar. 2021.

\bibitem{Liu2006_TRO}
Y.-H. Liu, H.~Wang, C.~Wang, and K.~K. Lam.
\newblock Uncalibrated visual servoing of robots using a depth-independent
  interaction matrix.
\newblock {\em IEEE Transactions on Robotics}, 22(4):804--817, Aug. 2006.

\bibitem{Lozano2000_Book}
R.~Lozano, B.~Brogliato, O.~Egeland, and B.~Maschke.
\newblock {\em Dissipative Systems Analysis and Control: Theory and
  Applications}.
\newblock Springer-Verlag, London, U.K., 2000.

\bibitem{Middleton1988_SCL}
R.~H. Middleton and G.~C. Goodwin.
\newblock Adaptive computed torque control for rigid link manipulators.
\newblock {\em Systems \& Control Letters}, 10(1):9--16, Jan. 1988.

\bibitem{Ortega1989_AUT}
R.~Ortega and M.~W. Spong.
\newblock Adaptive motion control of rigid robots: {A} tutorial.
\newblock {\em Automatica}, 25(6):877--888, Nov. 1989.

\bibitem{Seo2016_CDC}
D.~Seo.
\newblock Adaptive control for robot manipulator with guaranteed transient
  performance.
\newblock In {\em Proceedings of the IEEE Conference on Decision and Control},
  pages 2109--2114, Las Vegas, NV, USA, 2016.

\bibitem{Seo2009_SCL}
D.~Seo and M.~R. Akella.
\newblock Non-certainty equivalent adaptive control for robot manipulator
  systems.
\newblock {\em Systems \& Control Letters}, 58(4):304--308, Apr. 2009.

\bibitem{Sepulchre1996_IFAC}
R.~Sepulchre, M.~Jankovi{\'c}, and P.~V. Kokotovi{\'c}.
\newblock Integrator forwarding: a new recursive nonlinear robust design.
\newblock In {\em IFAC World Congress}, pages 1966--1971, San Francisco, CA,
  USA, 1996.

\bibitem{Slotine1988_CSM}
J.-J.~E. Slotine.
\newblock Putting physics in control--{T}he example of robotics.
\newblock {\em IEEE Control Systems Magazine}, pages 12--17, December 1988.

\bibitem{Slotine1990_TAC}
J.-J.~E. Slotine and M.~D.~D. Benedetto.
\newblock Hamiltonian adaptive control of spacecraft.
\newblock {\em IEEE Transactions on Automatic Control}, 35(7):848--852, Jul.
  1990.

\bibitem{Slotine1987_IJRR}
J.-J.~E. Slotine and W.~Li.
\newblock On the adaptive control of robot manipulators.
\newblock {\em The International Journal of Robotics Research}, 6(3):49--59,
  Sep. 1987.

\bibitem{Slotine1989_AUT}
J.-J.~E. Slotine and W.~Li.
\newblock Composite adaptive control of robot manipulators.
\newblock {\em Automatica}, 25(4):509--519, Jul. 1989.

\bibitem{Slotine1991_Book}
J.-J.~E. Slotine and W.~Li.
\newblock {\em Applied Nonlinear Control}.
\newblock Prentice-Hall, Englewood Cliffs, NJ, 1991.

\bibitem{Spong2006_Book}
M.~W. Spong, S.~Hutchinson, and M.~Vidyasagar.
\newblock {\em Robot Modeling and Control}.
\newblock Wiley, Hoboken, NJ, 2006.

\bibitem{Spong1990_TAC}
M.~W. Spong and R.~Ortega.
\newblock On adaptive inverse dynamics control of rigid robots.
\newblock {\em IEEE Transactions on Automatic Control}, 35(1):92--95, Jan.
  1990.

\bibitem{Takegaki1981_ASME}
M.~Takegaki and S.~Arimoto.
\newblock A new feedback method for dynamic control of manipulators.
\newblock {\em Journal of Dynamic Systems, Measurement, and Control},
  103(2):119--125, Jun. 1981.

\bibitem{Teel1992_IFAC}
A.~R. Teel.
\newblock Using saturation to stabilize a class of single-input partially
  linear composite systems.
\newblock In {\em IFAC Nonlinear Control Systems Design}, pages 379--384,
  Bordeaux, France, 1992.

\bibitem{Wang2015_AUT}
H.~Wang.
\newblock Adaptive visual tracking for robotic systems without image-space
  velocity measurement.
\newblock {\em Automatica}, 55:294--301, May 2015.

\bibitem{Wang2017_TAC}
H.~Wang.
\newblock Adaptive control of robot manipulators with uncertain kinematics and
  dynamics.
\newblock {\em IEEE Transactions on Automatic Control}, 62(2):948--954, Feb.
  2017.

\bibitem{Wang2020_AUT}
H.~Wang.
\newblock Differential-cascade framework for consensus of networked
  {L}agrangian systems.
\newblock {\em Automatica}, 112:108620, Feb. 2020.

\bibitem{Wang2020b_AUT}
H.~Wang.
\newblock Towards manipulability of interactive {L}agrangian systems.
\newblock {\em Automatica}, 119:108913, Sep. 2020.

\bibitem{Wang2019_ACC}
H.~Wang, W.~Ren, and C.~C. Cheah.
\newblock Forwardstepping: {A} new approach for control of dynamical systems.
\newblock In {\em Proceedings of the American Control Conference}, pages
  1208--1215, Philadelphia, PA, USA, 2019.

\bibitem{Wang2020_arXiv}
H.~Wang, W.~Ren, and C.~C. Cheah.
\newblock A differential-cascaded paradigm for control of nonlinear systems.
\newblock {\em arXiv preprint arXiv:2012.14251}, Dec. 2020.

\bibitem{Xia2020_IJACSP}
D.~Xia, X.~Xue, H.~Wen, and L.~Li.
\newblock Immersion and invariance adaptive tracking control for robot
  manipulators with a novel modified scaling factor design.
\newblock {\em International Journal of Adaptive Control and Signal
  Processing}, 34(1):110--125, Jan. 2020.

\bibitem{Yang2017_JGCD}
S.~Yang, M.~Akella, and F.~Mazenc.
\newblock Dynamically scaled {Immersion} and {Invariance} adaptive control for
  {Euler-Lagrange} mechanical systems.
\newblock {\em Journal of Guidance, Control, and Dynamics}, 40(11):2844--2856,
  Nov. 2017.

\end{thebibliography}

\end{document}